\documentclass[aps,prd,showpacs,twocolumn]{revtex4}

\usepackage{epsfig}
\usepackage{longtable}

\begin{document}

\title{Comment on the Nature of the $D_{s1}^*(2710)$ and $D_{sJ}^*(2860)$ Mesons} 
\author{ Stephen Godfrey\footnote{Email: godfrey@physics.carleton.ca} and Ian T. Jardine
\footnote{Present address:  Department of Physics, University of Toronto, M5S 1A7 Canada}}
\affiliation{
Ottawa-Carleton Institute for Physics, 
Department of Physics, Carleton University, Ottawa, Canada K1S 5B6 
}

\date{\today}

\begin{abstract}
Two charm-strange mesons, the $D_{s1}^*(2710)$ and the $D_{sJ}^*(2860)$, have recently been observed
by several experiments.  There has been speculation in the literature that the $D_{s1}^*(2710)$
is the $2^3S_1(c\bar{s})$ state and the $D_{sJ}^*(2860)$ is the $1^3D_1(c\bar{s})$ state.  
In this paper we explore this and other explanations 
in the context of the relativized quark model and the pseudoscalar emission decay model.
We conclude that the $D_{s1}^*(2710)$ is most likely the $1^3D_1 (c\bar{s})$ state and the 
$D_{sJ}^*(2860)$ is most likely the $1^3D_3 (c\bar{s})$ state with the $1D_2$ resonances also
contributing to the observed signals and explaining the observed ratios of branching ratios to $D^*K$ and $DK$ 
final states. We point out that measuring the $D_{sJ}^*(2860)$ spin can support or eliminate this
explanation and that there are six excited $D_s$ states in this mass region; the $2^3S_1$, $2^1S_0$,
$1^3D_1$, $1^3D_3$ and two $1D_2$ states.  Observing some of the missing states would help confirm
the nature of the $D_{s1}^*(2710)$ and the $D_{sJ}^*(2860)$ states.
\end{abstract}

\maketitle

\section{Introduction}

Heavy-light mesons provide a unique window into heavy quark dynamics and therefore provide an important
test of our understanding of quantum chromodynamics in the non-perturbative regime 
\cite{Rosner:1985dx,Eichten:1993ub,Swanson:2006st}.  
In recent years three new 
excited charm-strange mesons have been observed for the first time which can test calculations and
help improve our understanding of hadron spectroscopy.  They are 
the $D_{s1}^*(2700)^\pm$ \cite{Aubert:2006mh,Brodzicka:2007aa,Aubert:2009ah,Aaij:2012pc},
$D_{sJ}^*(2860)^\pm$ \cite{Aubert:2006mh,Aubert:2009ah,Aaij:2012pc}, 
and $D_{sJ}^*(3040)^+$ \cite{Aubert:2009ah}. 
We will focus on the first two states which have been observed by multiple experiments.
The Particle Data Group averages for the masses, decay widths and ratios of branching 
fractions for the $D_{s1}^*(2700)^\pm$ and $D_{sJ}^*(2860)^\pm$ 
are \cite{Beringer:1900zz}:
\begin{equation}
M(D^*_{s1}(2710)^\pm ) = 2709\pm 4 \; \hbox{MeV} 
\end{equation}
\begin{equation}
\Gamma(D^*_{s1}(2710)^\pm ) = 117 \pm 13 \; \hbox{MeV}
\end{equation}
\begin{equation}
{{\Gamma (D^*_{s1} \to D^* K)} \over {\Gamma (D^*_{s1} \to D K)}} 
= 0.91 \pm 0.13 (\hbox{stat}) \pm 0.12 (\hbox{syst})
\end{equation}
and
\begin{equation}
M(D^*_{sJ}(2860)^\pm ) = 2863^{+4.0}_{-2.6} \; \hbox{MeV} 
\end{equation}
\begin{equation}
\Gamma(D^*_{sJ}(2860)^\pm ) = 58 \pm 11 \; \hbox{MeV}
\end{equation}
\begin{equation}
{{\Gamma (D^*_{sJ} \to D^* K)} \over {\Gamma (D^*_{sJ} \to D K)}} 
= 1.10\pm 0.15 (\hbox{stat}) \pm 0.19 (\hbox{syst})
\end{equation}
Both states are observed decaying into both $DK$ and $D^*K$ so have natural
parity $J^P= 1^-$, $2^+$, $3^-, \ldots$. 
In this paper we compare the observed properties of these states to the mass predictions of the relativized quark 
model and decay predictions of a pseudoscalar emission model \cite{godfrey85xj} to determine their spectroscopic
assignments. This leads to further predictions that can test these assignments and help fill gaps in 
excited $D_s$ multiplets.

The  $D_{s1}^*(2710)^\pm$ and $D_{sJ}^*(2860)^\pm$ mesons have been studied in the context of various models
\cite{Close:2006gr,Zhang:2006yj,Colangelo:2006rq,Colangelo:2007ds,vanBeveren:2006st,vanBeveren:2009jq,Zhang:2009nu,Ebert:2009ua,Zhong:2008kd,Zhong:2009sk,Chen:2009zt,Chen:2011rr,Li:2009qu,Li:2010vx,Badalian:2011tb,Yuan:2012ej,Guo:2011dd,Wang:2013mml}.  
The $D_{s1}^*(2710)^\pm$
has been identified with the first radial excitation of the $D_{s1}^*(2112)^\pm$ 
or the  $D_{s}^*(1^3D_1)$ or some mixture of them 
\cite{Close:2006gr,Zhang:2006yj,Colangelo:2007ds,Zhong:2009sk,Zhang:2009nu,Chen:2009zt,Li:2009qu,Ebert:2009ua,Li:2010vx,Chen:2011rr,Wang:2013mml,Yuan:2012ej} 
and the $D_{sJ}^*(2860)^\pm$  
as the $D_s(2^3P_0)$ \cite{Close:2006gr,vanBeveren:2006st},
the $D_{s}^*(1^3D_1)$ or the  $D_{s}^*(1^3D_3)$ 
\cite{Colangelo:2006rq,Zhang:2006yj,Zhong:2009sk,Zhang:2009nu,Chen:2009zt,Li:2009qu,Li:2010vx,Chen:2011rr,Yuan:2012ej}.  
The $D_s(2^3P_0)$ does not appear to be a viable explanation 
because the $D_{sJ}^*(2860)^\pm$ has been observed in
both  $DK$ and $D^*K$ final states, although van Beveren and Rupp \cite{vanBeveren:2009jq}
argue that the signal could be the result of two overlapping resonances.
The 
theoretical predictions for these states are not completely consistent with their observed properties
so it is  useful to further test calculations against the experimental measurements. 
In the past we have found that the pseudoscalar emmission model \cite{godfrey85xj}
for OZI allowed hadronic decays
provides a useful consistency check for other models such as the $^3P_0$ strong decay model 
\cite{Godfrey:1986wj,Blundell:1995ev,Blundell:1995au}.  We therefore calculate the strong decay widths
of excited charm-strange mesons using the pseudoscalar emission model in this spirit, that it is useful
to compare the predictions of different models while acknowledging that some other models can be applied
to a broader range of decays than the pseudoscalar emission model.

In this paper we study the decay widths of excited $D_s$ mesons and compare the predicted and
measured properties of the $D_{s1}^*(2710)^\pm$ and  $D_{sJ}^*(2860)^\pm$  states.  
In the following section we begin by giving the relativized quark model mass predictions 
for the charm-strange mesons \cite{godfrey85xj}.  
We then give the partial decay widths for the $2S$ and $1D$ 
charm-strange mesons calculated using the pseudoscalar emmission model.  In section III we use these results to
discuss the possible spectroscopic assignments of the $D_{s1}^*(2710)^\pm$ and  $D_{sJ}^*(2860)^\pm$  states.
We summarize our conclusions in section IV.

\begin{figure}[t]
\begin{center}
\centerline{\epsfig{file=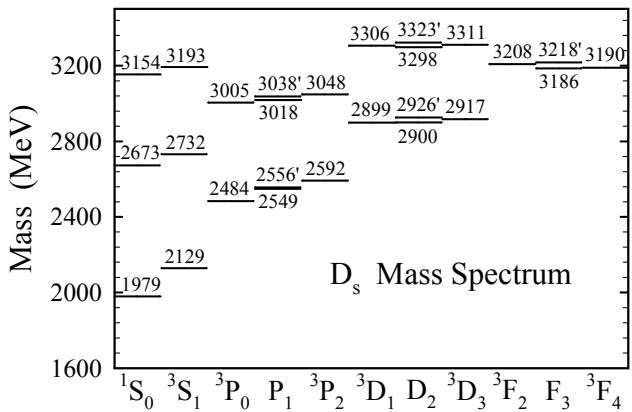,width=3.3in,clip=}}
\end{center}
\caption{The $D_s$ mass spectrum. The mixed $L_{J=L}$ states are described by eqn.~\ref{eqn:mixing}
with the primed states in the figure corresponding to the primed mixed state in 
eqn.~\ref{eqn:mixing} and the following mixing angles:  $\theta_{1P}^{c\bar{s}}=-37.5^\circ$, 
$\theta_{2P}^{c\bar{s}}=-30.4^\circ$, $\theta_{1D}^{c\bar{s}}=-38.5^\circ$, 
$\theta_{2P}^{c\bar{s}}=-37.7^\circ$ and $\theta_{1F}^{c\bar{s}}=-39.3^\circ$.
}
\label{Fig1}
\end{figure}

\section{$2S$ and $1D$ $D_s$ Properties}

\subsection{Spectroscopy}

We start with the mass predictions for the charm-strange mesons of the relativized quark model.
The details of this model can be found in Ref.~\cite{godfrey85xj} 
and \cite{Godfrey:1985by,godfrey85b,Godfrey:2004ya,Godfrey:2005ww} 
and we do not repeat them here.
This model has ingredients common to many quark potential models
\cite{Radford:2009bs,Ebert:1997nk,Ebert:2009ua,zeng95,Gupta:1994mw}.
Almost all such models are based on some variant of the Coulomb plus 
linear potential expected from QCD
and relativistic effects are often included at some level.  
The relativized quark 
model has been reasonably successful in describing most known mesons. 
However in recent years, starting with the discovery of the 
$D_{sJ}(2317)$ \cite{Aubert:2003fg,Besson:2003cp,Krokovny:2003zq} and $X(3872)$ states \cite{Choi:2003ue}, 
an increasing number of states have been observed that do not fit into this picture 
\cite{Godfrey:2008nc,Godfrey:2009qe,Braaten:2013oba,DeFazio:2012sg}  
pointing to the need to include physics which has hitherto been neglected such as 
coupled channel effects \cite{Eichten:2004uh}.  As a consequence of neglecting coupled channel effects
and the crudeness of the relativization procedure we do not 
expect the mass predictions to be accurate to better than $\sim 10-20$~MeV.
The mass predictions for this model are shown in Fig.~\ref{Fig1}.
Lattice gauge theory predictions \cite{Moir:2013ub} 
for the excited $1^-$ and $3^-$ charm-strange mesons are 
given in Table~\ref{tab:models} for comparison.  
See also Ref.~\cite{Mohler:2011ke} and \cite{Bali:2011dc}.

For the case of a quark and antiquark of unequal mass, charge conjugation
parity is no longer a good quantum number so that states with different 
total spins but with the same total angular momentum, such as
$^3P_1 -^1P_1$ and $^3D_2 -^1D_2$ pairs, can mix via
the spin orbit interaction or some other mechanism such as mixing via coupled channels.
Consequently, the physical $J=2$ $D$-wave states are linear
combinations of $^3D_2$ and $^1D_2$ which we describe by:
\begin{eqnarray}
\label{eqn:mixing}
D_{2}  & = {^1D_2} \cos\theta_{nD} + {^3D_2} \sin\theta_{nD} \nonumber \\   
D_{2}^{\prime} & =-{^1D_2} \sin\theta_{nD} + {^3D_2} \cos \theta_{nD} 
\end{eqnarray}
where $D\equiv L=2$ designates the relative angular momentum of the $q\bar{q}$ 
pair and the subscript $J=2$ is the total angular momentum of the $q\bar{q}$ 
pair which is equal to $L$ with analogous expressions for other values of $L$.  
We obtain $\theta_{1D}=-38.5^\circ$ (for $c\bar{s}$).
This notation implicitly implies $L-S$ 
coupling between the quark spins and the relative orbital angular momentum.  
In the heavy quark limit (HQL) in which the heavy quark mass $m_Q\to \infty$, 
the states can be described by the total angular momentum of the
light quark, $j_q$, which couples to the spin of the heavy quark and
corresponds to $j-j$ coupling. In this limit the state that is mainly spin singlet has $j_q=l+{1\over 2}$
while the state that is mainly spin triplet has $j_q=l-{1\over 2}$ and is labelled with a prime \cite{Cahn:2003cw}.
For $L=2$ the HQL gives 
rise to two doublets,  $j_q=3/2$ and $j_q=5/2$ with $\theta_{D}=-\tan^{-1}(\sqrt{2/3})=-39.2^\circ$ where the
minus sign arises from our $c\bar{s}$ convention
\cite{Cahn:2003cw,Zhang:2009nu,DiPierro:2001uu,Colangelo:2006rq,Colangelo:2007ds,Chen:2009zt,Chen:2011rr}.  
In some approaches it is more natural to use 
 the $j-j$ basis but because we solve our 
Hamiltonian equations assuming $L-S$ eigenstates and then include the 
$LS$ mixing we use the notation of eqn. \ref{eqn:mixing}.  
It is straightforward to
transform between the $L-S$ basis and the $j-j$ basis \cite{Cahn:2003cw}.  
We also note that the 
definition of the mixing angles is frought with ambiguities and
one should be extremely careful comparing predictions from different 
publications \cite{barnes}.

\subsection{Strong Decays}

We calculate decay widths using the pseudoscalar emission model 
(see Ref.~\cite{godfrey85xj} and references therein).
There are a number of predictions for $D_s$ decay widths in the literature using the $^3P_0$ model 
\cite{Close:2005se,Close:2006gr,Zhang:2006yj,Li:2009qu,Li:2010vx,Yuan:2012ej}
and other models \cite{Colangelo:2006rq,Colangelo:2007ds,Zhong:2009sk,DiPierro:2001uu,Chen:2011rr}.
A weakness of the pseudoscalar emission model is that it can only calculate partial widths that include 
light (non-charm) pseudoscalar mesons in the final state.  
For a few states that we consider, decays to light vector mesons are kinematically allowed.  
We expect the partial widths to a charm plus light vector to be small based on limited phase space. 
While this is generally supported by other calculations, in a few cases 
the BR to these final states was found to be as large as $\sim 20\%$.  
The $^3P_0$ model is more general and can therefore also be used in these cases.
Nevertheless we have found that the pseudoscalar emission model provides a useful
check of those results \cite{Godfrey:1986wj,Blundell:1995ev,Blundell:1995au}.  
In this spirit we calculate strong decays of excited $D_s$ mesons using the pseudoscalar 
emission model as another approach to
understand the nature of the recently observed $D_s$ states.  In this model the decay is assumed
to proceed through a single-quark transition. While the details of this model are given elsewhere,
for completeness,  we give the amplitude \cite{godfrey85xj}:
\begin{widetext}
\begin{eqnarray}
A_{q(\bar{q})} \left( M^*(\vec{k},s)  \to M(\vec{k}',s) P^i(\vec{q}) \right)  &  \nonumber \\
 =\pm i { { \sqrt{2\omega 2\omega'} } \over {(2\pi)^{9/2}} } 
\langle M(s') |  &  
\left\{  \left( { g + {h \over 2} { {m_{q(\bar{q})} - m_{\bar{q}(q)}} \over {m_q+m_{\bar{q}} } } }\right)
\vec{\sigma}_{q(\bar{q})} \cdot \vec{q} 
\pm  h \vec{\sigma}_{q(\bar{q})} \cdot \overleftarrow{p}' \right\}
e^{\mp i {m_{\bar{q}(q)} \over {m_q+m_{\bar{q}} } } \vec{q}\cdot \vec{r}} X^i_{q(\bar{q})} | M^*(s) \rangle
\end{eqnarray}
\end{widetext}
where $\vec{\sigma}_{q(\bar{q})}/2$ and $\vec{r}_{q(\bar{q})}$ are the spin and position of the quark (antiquark),
$\overleftarrow{p}'= -i \overleftarrow{\nabla}$ is the momentum operator 
acting on the final-state wave function and 
the upper (lower) sign refers to the $q (\bar{q})$ case.  The $X^i_{q(\bar{q})}$ are flavour operators.  
The calculations are most readily performed by taking $\vec{q}=q\hat{z}$ thereby calculating helicity amplitudes
$H_m$ where $m=s'=s$ and then tranforming to the usual partial-wave basis.  The details of this approach are
given in Ref.~\cite{godfrey85xj}. 

In our calculations we use harmonic oscillator wave functions with the oscillator parameter 
obtained by fitting the rms radius of the harmonic oscillator wavefunction to the 
rms radius of the full wavefunction.
Because the harmonic oscillator parameter, $\beta$, enters the decay expressions in combinations, we averaged 
the initial and final state $\beta$'s
to obtain the effective $\beta_{c\bar{s}}$ used in our numerical results.
This allows us to calculate the amplitudes analytically, revealing
relations between them that lets us classify the resulting amplitudes into two
classes: ``structure independent'' amplitudes which have only momentum dependence dictated by angular momentum
considerations along with the elastic form factor $F(q^2)$ defined below, and 
``structure dependent'' amplitudes which 
have additional polynomial momentum dependences which are sensitive to the 
structure of the states.  

The resulting partial wave amplitudes are given in Table~\ref{tab:decays} with the
following definitions for the amplitudes appearing in the table.
The following form factor is
factored out and needs to be included to obtain numerical values:
\begin{equation}
F(q^2)=\left( { q\over{2\pi} } \right)^{1/2} \exp \left[ { - {1\over 4}
\left( { m_c \over {m_c+m_{\bar{s}} } } \right)^2 {q^2\over{\beta_{c\bar{s}}^2}}
} \right]
\end{equation}
The amplitudes in Table~\ref{tab:decays} are the ``structure independent'' amplitude
\begin{equation}
A_{c\bar{s}}= \left[ { g +{h\over 2} \left( { m_{\bar{s}} \over {m_c+m_{\bar{s}} } } \right) } \right] \beta
\end{equation}
and the ``structure dependent'' amplitudes
\begin{equation}
P_{c\bar{s}}= \left\{ { 3h - {3\over 2} A_{c\bar{s}} \left( { m_c \over {m_c+m_{\bar{s}} } } \right) 
{{q^2}\over {\beta\beta_{c\bar{s}}^2} } }\right\}  \beta_{c\bar{s}}
\end{equation}
\begin{equation}
\label{eqn:d_decay}D_{c\bar{s}}= \left\{ { 3h - {3\over 5} A_{c\bar{s}} \left( { m_c \over {m_c+m_{\bar{s}} }} \right) 
{{q^2}\over {\beta\beta_{c\bar{s}}^2} } 
}\right\}  \beta_{c\bar{s}}
\end{equation}
In these expressions $m_s=0.5$~GeV and $m_c=1.7$~GeV are the relevant constituent 
quark masses used in the decay calculation, 
$\beta=0.4$~GeV and $\beta_{c\bar{s}}=0.53$ are harmonic oscillator wavefunction 
parameters used in obtaining these amplitudes.  $\beta$ is 
taken from the light meson decay analysis of Ref.~\cite{godfrey85xj} and 
$\beta_{c\bar{s}}$ was obtained as described above.
Rather than calculate these various reduced amplitudes in terms of $g$ and $h$
we approximate the two amplitudes in terms of two parameters, $A_{c\bar{s}}$ 
and $S_{c\bar{s}}$ and use the numerical values 
$S_{c\bar{s}}=3.27$ and $A_{c\bar{s}}=1.67$ obtained from the light meson decay analysis of Ref.~\cite{godfrey85xj}.
Further, following Ref.~\cite{godfrey85xj}, we take $P_{c\bar{s}}=D_{c\bar{s}}=S_{c\bar{s}}=3.27$.
We found that if we instead use
fitted values for $g$ and $h$ we obtain very similar numerical results for the partial widths.

To simplify the notation in the table we define the amplitudes:
\begin{equation}
\tilde{A}_{c\bar{s}} = A_{c\bar{s}} \left( { m_c\over{m_c + m_{\bar{s}} } }\right)^2 
{q^3\over{\beta \beta_{c\bar{s}}^2 } } F(q^2)
\end{equation}

\begin{equation}
\tilde{P}_{c\bar{s}} = P_{c\bar{s}} \left( { m_c\over{m_c + m_{\bar{s}} } }\right)
{q\over{ \beta_{c\bar{s}} } } F(q^2)
\end{equation}

\begin{equation}
\tilde{D}_{c\bar{s}} = D_{c\bar{s}} \left( { m_c\over{m_c + m_{\bar{s}} } }\right)
{q\over{\beta_{c\bar{s}} } } F(q^2)
\end{equation}

The amplitudes and partial widths for the $2S$ and $1D$ multiplets are given in Table~\ref{tab:decays}. 
The widths in column 4 of Table~\ref{tab:decays} were calculated using the predicted masses for the initial states 
and the measured  values from the PDG \cite{Beringer:1900zz} for the final states.  
The widths shown in column 5 were obtained using the measured masses of 
the $D_{s1}^*$ and $D_{sJ}^*$ states:
$M(2^3S_1)=M(D_{s1}^*)=2709$~MeV and  $M(1^3D_1)=M(1^3D_3)=M(D_{sJ}^*)=2863$~MeV.
In Table~\ref{tab:models} we compare our results to other calcuations where the widths shown here 
used the observed $D_{s1}^*$ and $D_{sJ}^*$ masses.
For the $D$-wave states we find that there are two narrow and two broad states corresponding to the 
$j=5/2$ and $3/2$ doublets of the  $m_Q\to \infty$ limit which are 
composed of the $(3^-, 2^-)$ and $(2^-, 1^-)$ states respectively. The $j=5/2$ states
are narrower due to the higher angular momentum barrier.  This pattern is similar to what is expected and 
observed for the $P$-waves states \cite{Rosner:1985dx,Godfrey:1986wj,Godfrey:2005ww}.  
This may be an important piece of the puzzle which we will return to in the next section.

\begin{table*}[t]
\caption{Partial widths for the $2S$ and $1D$ $c\bar{s}$ mesons calculated using the pseudoscalar emmision 
model.  The widths in column 4 were calculated using the predicted masses for the initial states 
and the PDG values \cite{Beringer:1900zz} for the final states.
The widths given in the last column were calculated using $M= 2710$~MeV for
the $2^3S_1$ initial state and $M= 2860$~MeV for the $1^3D_1$ and $1^3D_3$ initial states. The angles
$\theta_{c\bar{s}}=-38.5^\circ$ and $\phi= \tan^{-1}(\sqrt{2/3}) \simeq 39.23^\circ$. Details of the 
calculations are given in the text.
\label{tab:decays}}
\begin{tabular}{lllcc} \hline \hline
State & Decay  & Amplitude & Width & Width \\
	&	&	& (MeV) & (MeV) \\
\hline 
$D_s^*(2^3S_1)(2732)$ 
	& $D_s^*(2^3S_1)\to DK$ & $-i \sqrt{2\over{81}} \; \widetilde{P}_{c\bar{s}} $ & 13 & 12 \\
	& $D_s^*(2^3S_1)\to D^*K$ & $ -i \sqrt{4\over{81}} \; \widetilde{P}_{c\bar{s}} $ & 13 & 12 \\
	& $D_s^*(2^3S_1)\to D_s\eta$ & $ +i \sqrt{1\over{162}} \; \widetilde{P}_{c\bar{s}} $ & 1.7 & 1.4 \\
	& $D_s^*(2^3S_1)\to D_s^*\eta$ & $+i \sqrt{1\over{81}} \; \widetilde{P}_{c\bar{s}} $ & 0.7 & 0.4 \\
\hline
$D_s(2^1S_0)(2673)$ 
	& $D_s(2^1S_0)\to D^*K$ & $ +i \sqrt{2\over{27}} \; \widetilde{P}_{c\bar{s}}$ & 13 &  \\
	& $D_s(2^1S_0)\to D_s^*\eta$ & $ -i \sqrt{1\over{54}} \; \widetilde{P}_{c\bar{s}} $ & 0.1 & \\
\hline
$D_s^*(1^3D_3)(2917)$
	& $D_s^*(1^3D_3)\to DK$ 
		& $+i \sqrt{1\over{35}} \; \widetilde{A}_{c\bar{s}} $ & 27 & 19 \\
	& $D_s^*(1^3D_3)\to D^*K$ 
		& $+i \sqrt{4\over{105}} \; \widetilde{A}_{c\bar{s}} $ & 13 & 8\\
	& $D_s^*(1^3D_3)\to D_s\eta$ 
		& $-i \sqrt{1\over{140}} \; \widetilde{A}_{c\bar{s}} $ & 2.7 & 1.6 \\
	& $D_s^*(1^3D_3)\to D_s^*\eta$ 
		& $-i \sqrt{1\over{105}} \; \widetilde{A}_{c\bar{s}} $ & 0.8 & 0.3 \\
\hline
$D_s(D_{2}')(2926)$
	& ${D_s}_{2}'\to [D^*K]_P$ 
		& $-i \sqrt{5\over{27}} \; \widetilde{D}_{c\bar{s}} \cos(\theta+\phi)$ & 116 & \\
	& ${D_s}_{2}'\to [D^*K]_F$ 
		& $+i \sqrt{1\over{15}}  \; \widetilde{A}_{c\bar{s}} \sin(\theta+\phi)$ & $\sim 0$ & \\
	& ${D_s}_{2}'\to [D^*\eta]_P$ 
		& $+i \sqrt{5\over{108}} \; \widetilde{D}_{c\bar{s}} \cos(\theta+\phi)$ & 17 & \\
	& ${D_s}_{2}'\to [D^*\eta]_F$ 
		& $-i \sqrt{1\over{60}}  \; \widetilde{A}_{c\bar{s}} \sin(\theta+\phi)$ & $\sim 0$ & \\
\hline
$D_s(D_{2})(2900)$
	& ${D_s}_{2}\to [D^*K]_P$ 
		& $-i \sqrt{5\over{27}} \; \widetilde{D}_{c\bar{s}} \sin(\theta+\phi)$ 	& $\sim 0$ & \\
	& ${D_s}_{2}\to [D^*K]_F$ 
		& $-i \sqrt{1\over{15}} \; \widetilde{A}_{c\bar{s}} \cos(\theta+\phi)$ & 20 & \\
	& ${D_s}_{2}\to [D^*\eta]_P$ 
		& $+i \sqrt{5\over{108}} \; \widetilde{D}_{c\bar{s}} \sin(\theta+\phi)$ & $\sim 0$ & \\
	& ${D_s}_{2}\to [D^*\eta]_F$ 
		& $+i \sqrt{1\over{60}} \; \widetilde{A}_{c\bar{s}} \cos(\theta+\phi)$ 	& 1.0 & \\
\hline
$D_s^*(1^3D_1)(2899)$
	& $D_s^*(1^3D_1)\to DK$ & $+i \sqrt{10\over{81}} \; \widetilde{D}_{c\bar{s}} $ & 101 & 93 \\
	& $D_s^*(1^3D_1)\to D^*K$ & $-i \sqrt{5\over{81}} \; \widetilde{D}_{c\bar{s}} $ & 36 & 32 \\
	& $D_s^*(1^3D_1)\to D_s\eta$ & $ -i \sqrt{5\over{162}} \; \widetilde{D}_{c\bar{s}} $ & 18 & 16 \\
	& $D_s^*(1^3D_1)\to D_s^*\eta$ & $+i \sqrt{5\over{324}} \; \widetilde{D}_{c\bar{s}} $ & 5.0 & 4 \\
\hline
\hline
\end{tabular}
\end{table*}

Before proceeding to the discussion we make a brief digression regarding kinematically allowed decays 
to $DK^*$ final states.  As pointed out above, decays to light vector mesons are beyond the scope of the
pseudoscalar emission 
model.  However, we can use heavy quark symmetry to estimate their partial widths. In the heavy quark limit the
light quark angular momentum is separately conserved from the heavy quark spin. For the $1^3D_3$ state 
$j_q^P=5/2^-$ and for the $1^3D_1$ state $j_q^P=3/2^-$.  For the decay $1^3D_3\to DK$ the final state must
be in an $F$-wave to conserve angular momentum while for $1^3D_3\to DK^*$ the final state can be in a
$P$-wave.  Assuming the underlying amplitudes are similar and including angular momentum factors, elastic
form factors,  and the momentum factors appropriate to $P$ and $F$ waves we find
$\Gamma(1^3D_3\to DK^*) \simeq 0.08 \times \Gamma(1^3D_3\to DK)$ which is consistent with the various results
given in Table~\ref{tab:models}.  We make a similar estimate for the decays $1^3D_1\to DK$ and $1^3D_1\to DK^*$ 
which both have final states in $P$-waves.  For this case we estimate that the partial width to $DK^*$ 
is about 10\% of that to the $DK$ final state.  This is only consistent with one of the results listed in 
Table~\ref{tab:models} with the other two calculations predicting a signficantly larger partial width to  $DK^*$.
However those two cases have larger total widths so that in all cases the partial width to $DK^*$ will 
only modify the total width by at most $\sim 20\%$.

\begin{table*}
\caption{Comparison of the present calculation (PSEM) to other calculations in the literature. 
The errors for the Lattice results are statistical.  For our results, labelled PSEM, the masses listed
are from Ref.~\cite{godfrey85xj} but the partial widths are calculated using the measured masses.
For CS \cite{Close:2005se} the masses are from Ref.~\cite{godfrey85xj}
and not all calculated partial widths are listed but they are included in the total width.  
For ZLDZ \cite{Zhang:2006yj}, ZZ \cite{Zhong:2009sk}, LM  \cite{Li:2009qu} and YCZ \cite{Yuan:2012ej}
the masses are not predictions but are the measured masses used as input.
We do not list all small partial widths calculated by ZLDZ and LM  but do include them in the total width.
YCZ 1 uses different harmonic oscillator wavefunction parameters for different states while YCZ 2
uses the same harmonic oscillator wavefunction parameters for all states. 
\label{tab:models}}
\begin{tabular}{lcccccccccc} \hline
State 
	& Experiment\cite{Beringer:1900zz} 
	& Lattice\cite{Moir:2013ub} 
	& PSEM 
	& CS\cite{Close:2005se} 
	& ZLDZ\cite{Zhang:2006yj} 
	& ZZ\cite{Zhong:2009sk} 
	& LM\cite{Li:2009qu}  
	& YCZ 1\cite{Yuan:2012ej} 
	& YCZ 2\cite{Yuan:2012ej} 
\\
\hline
$M(2^3S_1)(2709) $ & $D_{s1}$ $2709\pm 4$  & $2757 \pm 6$ & 2732 & 2730 &  2715  & 2710 & 2710 & 2709 & 2709 \\
$\Gamma(\to DK)$ & &   & 12  & 17 & 3.2 & 11 & 4.4 & 9.4 & 32 \\
$\Gamma(\to D^*K)$ & &   & 12  & 81 & 27.2 & 18.1 & 34.9 & 41 & 85 \\
$\Gamma(\to D_s\eta)$ &  &    & 1.4  & 2.6 & 0.05 & 1.7 & 0.8  & 2.0 & 20 \\
$\Gamma(\to D_s^*\eta)$&  &  & 0.4  & 4.1 & 0.54 & 0.7  & 1.4 & 2.0 & 11 \\
$\Gamma_{Total}$ & $117\pm 13$  &   & 25  & 105 & 32 & 31 & 41.4 & 55 & 148 \\ 
$\Gamma(\to D^*K)/\Gamma(\to DK)$ & $ 0.91\pm 0.18$ &  & 0.99  &  4.8 & 8.5 & 1.65 & 7.93 & 4.4 & 2.67 \\
\hline
$M(1^3D_1)(2709)$ & $D_{s1}$ $2709\pm 4$ & & 2899 &  & 2715 & 2710 & 2710 & 2709 & 2709 \\
$\Gamma(\to DK)$ & & & 59 & & 49.4 & 149 & 87 & 94 & 90 \\
$\Gamma(\to D^*K)$ &  & & 14 & & 8 & 36 & 37 & 39 & 44 \\
$\Gamma(\to D_s\eta)$ &  & & 7.2 & & 13.2 & 14 & 13 & 17 & 30 \\
$\Gamma(\to D_s^*\eta)$ &  &  & 0.5 & & 2.4 & 0.9 & 1.5 & 2.0 & 3.3 \\
$\Gamma_{Total}$ & $117\pm 13$ & & 81  &  & 73 & 200 & 138 & 152 & 167 \\ 
$\Gamma(\to D^*K)/\Gamma(\to DK)$ & $ 0.91\pm 0.18$  & & 0.25 & & 0.16 & 0.24 & 0.43 & 0.42 & 0.48 \\
\hline
$M(1^3D_1)(2863)$ & $D_{sJ}$ $2863^{+4.0}_{-2.6}$   & $2888\pm 5$ & 2899 & 2900 & 2860 & & 2862 & & \\ 
$\Gamma(\to DK)$ & &   & 93 & 120 & 84 & & 63 & &  \\
$\Gamma(\to D^*K)$ & &   & 32  & 74 & 14 & & 38 & &    \\
$\Gamma(\to D_s\eta)$ & &    & 16 & 39 & 24 & & 20 & &  \\
$\Gamma(\to D_s^*\eta)$ & &    & 4.0 & 17 & 2 &  & 8.5 & &  \\
$\Gamma(\to DK^*)$ &  & & -- & 81 & 7.8 & & 38 & & \\
$\Gamma_{Total}$ & $58\pm 11$ &  & 145 & 331 & 132 &  & 168 & &  \\ 
$\Gamma(\to D^*K)/\Gamma(\to DK)$ & $ 1.10\pm 0.24$ &   & 0.34 & 0.62 & 0.17 & & 0.60 & &  \\
\hline
$M(1^3D_3)(2863)$ & $D_{sJ}$ $2863^{+4.0}_{-2.6}$   & $2942\pm 6$ & 2917 & 2920 & 2860 & 2860 & 2862 & 2862 & 2862 \\ 
$\Gamma(\to DK)$ & &   & 19 & 82 & 22 & 24 & 36 & 33 & 42 \\
$\Gamma(\to D^*K)$ & &   & 8.1 & 67 & 13 & 10 & 27 & 23 & 24  \\
$\Gamma(\to D_s\eta)$ & &  & 1.6  & 4.5 & 1.2 & 1.7 & 1.6 & 1.9 & 6.2 \\
$\Gamma(\to D_s^*\eta)$ & &  & 0.3  & 2.2 & 0.3 & 0.3 & 0.6 & 0.7 & 1.5  \\
$\Gamma(\to DK^*)$ & & & -- & 14 & 0.71 & 0.2 & 2.7 & 2.1 & 1.4 \\
$\Gamma_{Total}$ & $58\pm 11$ &   & 29 & 222 & 37 & 36 & 67 & 60 & 75 \\ 
$\Gamma(\to D^*K)/\Gamma(\to DK)$  & $ 1.10\pm 0.24$ &  & 0.43 & 0.82  & 0.59 & 0.40 & 0.75 & 0.69 & 0.56 \\
\hline
\end{tabular}
\end{table*}

\section{On the Nature of the $D_s(2700)$ and $D_s(2860)$ States}

With the mass and width predictions for excited $D_s$ states in hand we examine whether the $D_{s1}^*(2710)^\pm$
state can be identified with the $2^3S_1(c\bar{s})$ or $1^3D_1(c\bar{s})$ and the $D_{sJ}^*(2860)^\pm$ with
the $1^3D_1(c\bar{s})$ or $1^3D_3(c\bar{s})$ states. In what follows we will refer
to the decay results summarized  in Table~\ref{tab:models}.

The predicted mass for the $2^3S_1(c\bar{s})$ is 2732~MeV compared to the PDG \cite{Beringer:1900zz}
averaged mass for the $D_{s1}^*(2710)$ of 
$2709 \pm 4$~MeV and the predicted masses for the $1^3D_1(c\bar{s})$ and $1^3D_3(c\bar{s})$ 
states are 2899 and 2917~MeV
respectively, which are compared to the measured $D_{sJ}^*(2860)$ mass of $2863^{+4.0}_{-2.6}$~MeV.  The 
measured $D_{s1}^*(2710)$ mass is consistent with the predicted $2^3S_1$ mass within the accuracy of the model.
The measured $D_{sJ}^*(2860)$ mass is $\sim 36$~MeV lower than the predicted 
$1^3D_1(c\bar{s})$ mass and $\sim 54$~MeV lower
than the predicted $1^3D_3(c\bar{s})$ mass.  Depending on how well the predicted and measured decay widths
agree one could accept the discrepancies in the mass predictions as being within the uncertainties of the model.

The predicted total width for the $2^3S_1(c\bar{s})$ is 26~MeV versus the $D_{s1}^*(2710)^\pm$
measured width of $117\pm 13$~MeV and the predicted value for the ratio of partial widths is 
$\Gamma(D_{s1}^*(2^3S_1) \to D^*K)/\Gamma(D_{s1}^*(2^3S_1) \to DK)=0.99$ versus the observed value of
$0.91 \pm 0.13 (\hbox{stat}) \pm 0.12 (\hbox{syst})$.  While the 
$\Gamma(D^*K)/\Gamma(DK)$ ratio is in good agreement there is significant disagreement for the total widths
even if we accept that the predicted widths could be off by a factor of two.
Similarly, the measured $D_{sJ}^*(2860)^\pm$ width and $\Gamma(D^*K)/\Gamma(DK)$ ratio are
$58 \pm 11 \; \hbox{MeV}$ and $1.10\pm 0.15 (\hbox{stat}) \pm 0.19 (\hbox{syst})$ respectively  compared 
to the predicted values of 145~MeV and 0.34 for the $1^3D_1(c\bar{s})$ state.  
We conclude that the decay properties of the
$D_{s1}^*(2710)^\pm$ and the  $D_{sJ}^*(2860)^\pm$ are not consistent with those of the 
$2^3S_1(c\bar{s})$ and $1^3D_1(c\bar{s})$ states.  This conclusion is consistent with those of other
calculations although it should be noted that Ref.~\cite{Close:2005se} 
and \cite{Yuan:2012ej} (unequal $\beta$ case)
find the $2^3S_1(c\bar{s})$ total width to be consistent with the $D_{s1}^*(2710)^\pm$ width although the 
$\Gamma(D^*K)/\Gamma(DK)$ ratio is in substantial disagreement.

\begin{figure*}[t]
\begin{center}
\centerline{\epsfig{file=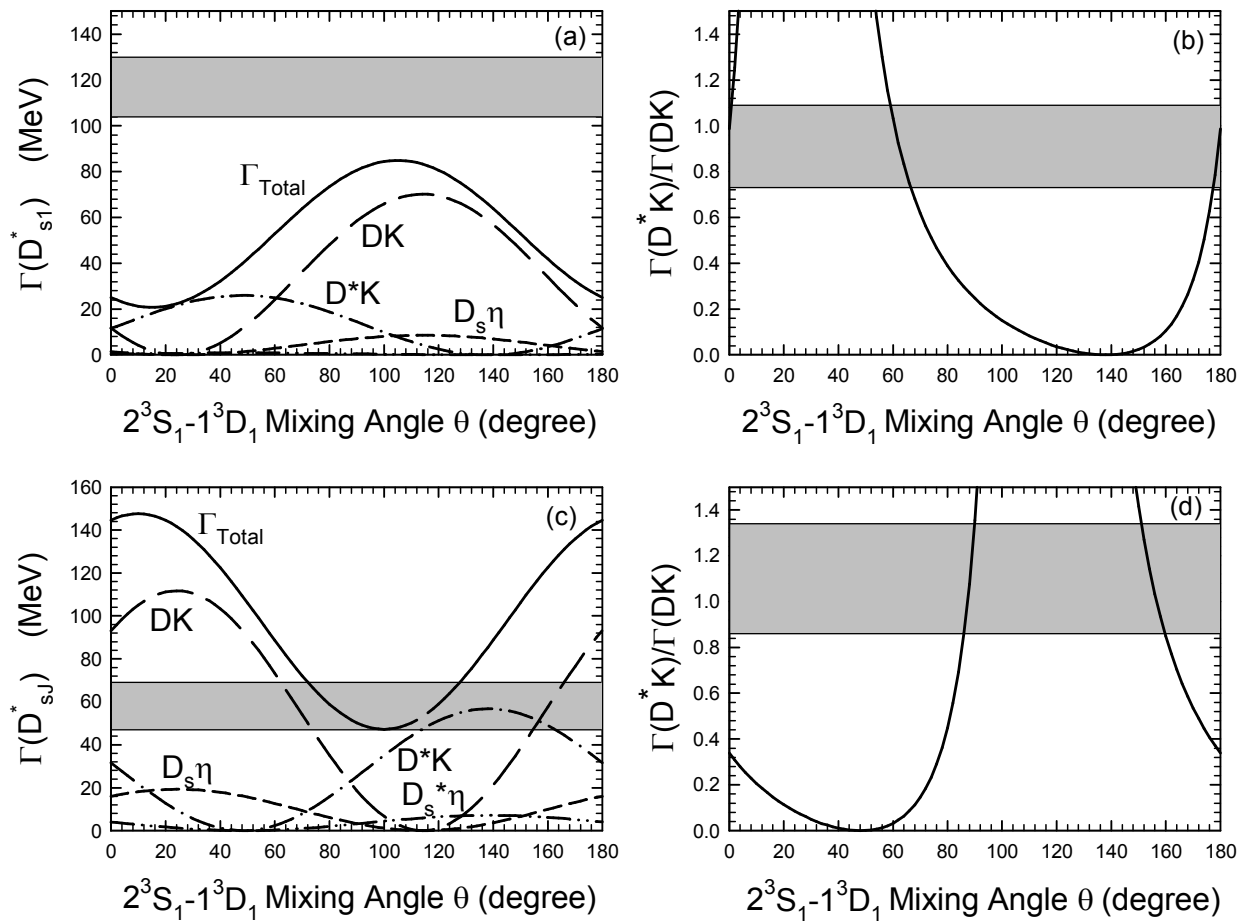,width=6.5in,clip=}}
\end{center}
\caption{Decay widths and the ratios $\Gamma(D^*K)/\Gamma(DK)$ of the $D_{s1}^*(2700)^\pm$
and the $D_{sJ}^*(2860)^\pm$ assuming they are linear combinations of the $2^3S_1$ and $1^3D_1$ $c\bar{s}$ 
states as a function of the $2^3S_1 -1^3D_1$ mixing angle defined in the text. The shaded regions are the 
$1-\sigma$ ranges for the experimental measurements.  Fig. (a) and (b) are for the $D_{s1}(2700)$
state and (c) and (d) for the $D_{s1}(2860)$ state.  In Fig. (a) and (c) the solid lines are for the
total widths,  the long-dashed line for $DK$, dot-dashed for $D^*K$, short-dashed for $D_s \eta$ and
dash-dot-dot for $D_s^* \eta$ final states.  (The $D_{s1}^*(2700) \to D_s^* \eta$ 
partial width is too small to show up in Fig. (a).) }
\label{Fig2}
\end{figure*}

To understand the $D_{s}^*$ states, other possibilities have been put forward.  In one scenario the
$D_{s1}^*(2710)^\pm$ is a mixture of  $2^3S_1(c\bar{s})$ and $1^3D_1(c\bar{s})$ 
\cite{Close:2006gr,Zhong:2009sk,Li:2009qu,Li:2010vx,Chen:2011rr,Yuan:2012ej,Wang:2013mml}
and the  $D_{sJ}^*(2860)$ is its orthogonal partner \cite{Li:2009qu,Li:2010vx,Yuan:2012ej}:
\begin{eqnarray}
\label{eqn:sd_mixing}
D_{s1}^* & = 2^3S_1 \cos\theta + 1^3D_1 \sin\theta \nonumber \\   
D_{sJ}^* & =-2^3S_1 \sin\theta + 1^3D_1 \cos \theta .
\end{eqnarray}
The partial and total widths and the $\Gamma(D^*K)/\Gamma(DK)$ ratios for the $D_{s1}^*$ and $D_{sJ}^*$ 
mixed states defined by Eqn.~\ref{eqn:sd_mixing} are plotted as a function of the mixing angle in 
Fig.~\ref{Fig2}.  
Eqn.~\ref{eqn:sd_mixing} is a simplification as in general the mixing angle should depend on energy and a
coupled channel analysis is required.  
However, this simplification is adequate 
for a preliminary study to determine whether mixing is a viable explanation for the observed decay properties 
and whether a more detailed analysis is warranted.
The shaded bands in these plots represent the one standard deviation regions of
the measured values for these quantities \cite{Beringer:1900zz}.  
One can see that reasonable agreement is obtained with a 
$2^3S_1- 1^3D_1$ mixing angle of $\sim 90^\circ$.  Calculating the $\chi^2$ for the four observables finds 
the best fit for $\theta \simeq 88^\circ$.  In other words we obtain a reasonable fit for the decay widths assuming that the
$D_{s1}^*(2710)^\pm$ is primarily the $1^3D_1(c\bar{s})$ state and the $D_{sJ}^*(2860)$ is primarily
the $2^3S_1(c\bar{s})$ state.  This, however, is inconsistent with the mass predictions and implies
that the $2^3S_1(c\bar{s})$ is more massive than the $1^3D_1(c\bar{s})$. We are aware that in this scenario the 
predicted $D_{s1}^*(2710)^\pm$ width is smaller than the observed width but
as stated previously, we consider the difference to be within the predictive power of the model.

A more serious discrepancy is that for the $1^3D_1(c\bar{s})$ state 
the predicted $D_{s1}^*(2700)^\pm$ $\Gamma(D^*K)/\Gamma(DK)$ ratio is about a 
factor of two smaller than what is observed.  
It is possible that various corrections could bring the predicted value for this ratio into closer agreement 
with experiment.  For example,  the amplitudes for the decays $1^3D_1(c\bar{s}) \to D^*K$ and $DK$ are the
``structure dependent'' type, Eqn.\ref{eqn:d_decay}, which we approximated with a constant.  In 
previous calculations we found that this approximation gave reasonably good agreement with experiment 
and other models \cite{Godfrey:1986wj,Blundell:1995au,Godfrey:2005ww}.  
However, if in Eqn.\ref{eqn:d_decay} one takes into account the smaller phase space of the decay to $D^*K$ 
compared to $DK$  we expect an enhancement in the $\Gamma(D^*K)/\Gamma(DK)$ ratio, although it is unlikely to
be sufficient to bring it into agreement with experiment.

This brings us back to a point mentioned previously, that the 
physical $J=2$ states are mixtures of $1^3D_2$ and 
$1^1D_2$ resulting in one narrow and one broad state approximately degenerate with the $1^3D_1$ and $1^3D_3$
states.  Perhaps the peak seen at 2710~MeV has contributions from two states, 
the $1^3D_1$ state and one of the $1D_2$ 
states.  Because the $1D_2$ states can only decay to $D^*K$ due to conservation of angular momentum and parity, 
the  bump seen in the $D^*K$ final state 
might have contributions from  both the $1^3D_1$ and the $1D_2$ states and be enhanced 
relative to what would be predicted from the $1^3D_1$ on its own, thereby explaining the observed
$\Gamma(D^*K)/\Gamma(DK)$ ratio. With sufficient statistics it should be possible to study the angular
distributions of the final states and test this possibility.
This explanation leaves  the $D_{sJ}^*(2860)$ unaccounted for. 

A possibility put forward in the literature for the $D_{sJ}^*(2860)$ is that it is the $1^3D_3(c\bar{s})$ state
\cite{Zhang:2006yj,Colangelo:2006rq,Zhong:2009sk,Zhang:2009nu,Chen:2009zt,Li:2009qu,Li:2010vx,Chen:2011rr,Yuan:2012ej}.  
The predicted total width and $\Gamma(D^*K)/\Gamma(DK)$ ratio  are 29~MeV and 0.42 
respectively.  As already stated, we can accept that our width predictions are off by a factor of two but 
we still need to understand the discrepancy in the ratio of BR's. Unlike the $1^3D_1$ decays, the amplitudes
for the the $1^3D_3 \to D^*K$ and $DK$ are ``structure independent'' amplitudes so we do not expect
the ratio of BR's to change.  
 A possible explanation is the same as we 
suggested for the $D_{s1}^*(2710)^\pm$ state, that the bump at 2860~MeV 
could be explained as two overlapping resonances, the $1^3D_3(c\bar{s})$ state and a $1D_2$ state 
 which would explain the discrepancy in the $\Gamma(D^*K)/\Gamma(DK)$ ratio \cite{Zhong:2009sk}.

Putting these pieces together we arrive at the possibility that the 
$D_{s1}^*(2710)^\pm$ is the $1^3D_1$ state and the $D_{sJ}(2860)$ is the $1^3D_3$ state with overlapping $1D_2$ 
states that modify the observed $\Gamma(D^*K)/\Gamma(DK)$ ratios.  
However this does lead to a number of questions.  
First, with this explanation the measured $1^3D_1-1^3D_3$ splitting, $\sim 154$~MeV, 
 would be much much larger than the predicted splitting of 18~MeV.  
For comparison, in the strange meson system,  the only light-heavy system 
for which there are candidate states for the $1^3D_1$ and $1^3D_3$, the splitting is $59 \pm 28$~MeV.
So although we might not want to rule out such a large splitting it is inconsistent with the predictions
of this model. Next, if we identify the $D_{s1}^*(2710)^\pm$ with the $1^3D_1$, where is
the $2^3S_1$ state? 
Again, comparing to the strange mesons where the $2^3S_1$ is predicted to lie 200~MeV below the 
$1^3D_1$ state. This is roughly comparable to the $1^3D_1-2^3S_1$ splitting of 168~MeV predicted 
for the charm-strange mesons.  
However, in the strange sector, the mesons 
identified with the $2^3S_1$ and $1^3D_1$ states have a measured mass difference of 303~MeV.  
So it is possible that the model gets this wrong and the $2^3S_1 (c\bar{s})$ state is also much lighter than 
predicted.  
We note that B. Zhang {\it et al.} \cite{Zhang:2006yj} also come to this conclusion by calculating decay widths
using the $^3P_0$ model 
and  A. Zhang \cite{Zhang:2009nu} and Colangelo {\it et al.} \cite{Colangelo:2006rq}
come to the same conclusion treating
the $D_{s1}^*(2710)^\pm$ and $D_{sJ}^*(2860)$ as $j_q^P ={3\over 2}^-$ and ${5\over 2}^-$ states in the 
heavy quark limit and calculating the partial widths using an effective Lagrangian approach. 

We have examined three possible interpretations of the $D_{s1}^*(2710)^\pm$ and 
$D_{sJ}^*(2860)$ mesons based on the masses predicted by the relativized quark model and strong decay
properties predicted by the pseudoscalar emission model (PSEM).  None of these possibilities is without
flaws and we conclude that the most likely explanation is that the $D_{s1}^*(2710)^\pm$ and 
$D_{sJ}^*(2860)$ mesons are the $1^3D_1$ and $1^3D_3$ $c\bar{s}$ states.  However the predicted
ratios of partial widths for these states to $D^*K$ and $DK$ final states are in sharp disagreement with the BaBar 
measurements \cite{Aubert:2009ah}.  We suggest that this might be due to the $1D_{2}$ states, 
which decay  to $D^*K$, overlapping with the $1^3D_1$ and $1^3D_3$ states thereby enhancing the $D^*K$ signal.
In Table~\ref{tab:models} we compare our predictions to a broad range of calculations and for the most part 
find similar results within the uncertainties of the models. The most noteworthy exceptions are the results of
Close and Swanson \cite{Close:2005se} who predict much larger total widths for the 
$2^3S_1$, $1^3D_1$ and $1^3D_3$ states.  In addition, all calculations other than ours find much larger
values for the $\Gamma(2^3S_1 \to D^*K)/\Gamma(2^3S_1 \to DK)$ ratio.   
As a consequence, further measurements will be needed
to confirm or rule out the various quark model assignments of these states.

The most useful measurement is to determine the spin of the $D_{sJ}^*(2860)$ which would settle whether
it is the $1^3D_3(c\bar{s})$ or one of the $J^P=1^-$ states.  With sufficient statistics one could also 
determine whether there is more than one state contributing to both the $D_{s1}^*(2710)^\pm$ and 
$D_{sJ}^*(2860)$ bumps.  This would be able to rule out or confirm our suggestion of overlapping resonances.

Six states are predicted to lie in this mass region, the $2^1S_0$, $2^3S_1$ and the 4 D-wave states,  but 
only two states have been observed.  Observing the 4 missing states would go a long way to shed light on 
these states.  We predict the $D_s(2^1S_0)$ mass to be 2673~MeV and to be relatively narrow 
decaying predominantly to the $D^*K$ final state.  It should therefore be possible to observe it in $D^*K$ 
by the experiments that have seen the  $D_{s1}^*(2710)^\pm$ and $D_{sJ}^*(2860)$ in this final state.  
If the $D_{s1}^*(2710)^\pm$ is in fact the $1^3D_1$ $c\bar{s}$  it should also be possible
to observe the  $2^3S_1$ $c\bar{s}$ in $DK$ and $D^*K$ final states.  If it isn't, as stated previously, 
it will be important to measure the $D_{sJ}^*(2860)$ spin to distinguish between the $1^3D_1$ and
$1^3D_3$.  Finally,  we expect the $1D_2$ states to be close in mass to the $1^3D_1$ and $1^3D_3$
states so that high statistics will be needed to see them as separate resonances or to be able to
perform an angular momentum analysis that can distinguish two overlapping states with different angular
momentum \cite{Zhong:2009sk}.  It is possible that the four missing states have not yet 
been observed because of insufficient statistics
or that the invariant mass spectra are dominated by the tails of the 
$D_{s2}^*(2573)$ and $D_{s1}(2536)$ resonances in this mass region.

\section{Summary}

In this paper we attempted to understand the nature of the $D_{s1}^*(2710)^\pm$ and $D_{sJ}^*(2860)$
states as charm-strange mesons using the relativized quark model and the pseudoscalar emission model for their decays. 
We considered a number of possibilities;  that the  $D_{s1}^*(2710)^\pm$ is the $2^3S_1 (c\bar{s})$ state,
that it is a linear combination of  $2^3S_1$ and $1^3D_1 (c\bar{s})$ and the $D_{sJ}^*(2860)$ is its 
orthogonal partner.  We came to the conclusion that they are best described
as the $1^3D_1$ and $1^3D_3$ $(c\bar{s})$ states.  However, the predicted 
ratios of partial widths for these states to $D^*K$ and $DK$ final states do not agree with the measured
ratios.  We suggest that this could be a result of overlapping $D_2$ states  which enhances 
the $D^*K$ final state.  A crucial measurement to distinguish these possibilities is the measurement of
the  $D_{sJ}^*(2860)$ spin and with sufficient statistics to determine whether the bumps at 2710~MeV and 2860~MeV 
contain two resonances with different spins.  We also point out that 6 states are expected in this mass
region so that 4 are still to be accounted for.  Finding them would fill in gaps in the excited $D_s$ spectrum
and confirm the nature of the $D_{s1}^*(2700)^\pm$ and $D_{sJ}^*(2860)$ states.

\acknowledgments

This research was supported in part 
the Natural Sciences and Engineering Research Council of Canada under grant number 121209-2009 SAPIN.


\end{document}